\documentclass[%
 reprint,
 amsmath,amssymb,
 aps, showpacs
]{revtex4-2}

\usepackage{graphicx}
\usepackage{dcolumn}
\usepackage{bm}
\usepackage{verbatim}
\usepackage{array}
\usepackage{amsmath}
\usepackage{braket}
\usepackage{lineno}
\usepackage[bookmarksnumbered, pdfstartview=FitH,colorlinks,urlcolor=blue, citecolor=blue,linkcolor=blue,]{hyperref}
\usepackage{setspace}
\usepackage{subfigure}

\begin{document}

\preprint{APS/123-QED}

\title{Probing hyperon electric dipole moments with a full angular analysis}

\author{Jinlin Fu$^{1}$}
\email{jinlin.fu@ucas.ac.cn}
\author{Hai-Bo Li$^{1,2}$}
\author{Jian-Peng Wang$^{3,4}$}
\email{wangjp20@lzu.edu.cn}
\author{Fu-Sheng Yu$^{3,4,5}$}
\author{Jianyu Zhang$^{1}$}
\email{zhangjianyu@ucas.ac.cn}
\affiliation{%
        $^{1}$School of Physical Sciences, University of Chinese Academy of Sciences, Beijing 100049, People's Republic of China\\
        $^{2}$Institute of High Energy Physics, Chinese Academy of Sciences, Beijing 100049, People's Republic of China\\
        $^{3}$MOE Frontiers Science Center for Rare Isotopes, Lanzhou University, Lanzhou 730000, People's Republic of China \\
        $^{4}$School of Nuclear Science and Technology, Lanzhou University, Lanzhou 730000, People's Republic of China\\
        $^{5}$Center for High Energy Physics, Peking University, Beijing 100871, People's Republic of China\\
}%

\date{\today}

\begin{abstract}
The electric dipole moment (EDM) of elementary particles, arising from flavor-diagonal $CP$ violation, serves as a powerful probe for new physics beyond the Standard Model and holds the potential to provide novel insights in unraveling the puzzle of the matter-dominated Universe.
Hyperon EDM is a largely unexplored territory.
In this paper, we present a comprehensive angular analysis that focuses on entangled hyperon-antihyperon pairs in $J/\psi$ decays for the indirect extraction of hyperon EDM. The statistical sensitivities are investigated for BESIII and the proposed Super Tau-Charm Facility (STCF). Leveraging the statistics from the BESIII experiment, the estimated sensitivity for $\Lambda$ EDM can reach an impressive level of $10^{-19}$ $e$ cm, achieving a 3-orders-of-magnitude improvement over the only existing measurement in a fixed-target experiment at Fermilab with similar statistics. The estimated sensitivities for the $\Sigma^+$, $\Xi^-$, and $\Xi^0$ hyperons at the same level of $10^{-19}$ $e$ cm will mark the first-ever achievement and the latter two will be the first exploration of hyperons with two strange valence quarks. The EDM measurements for hyperons conducted at the BESIII experiment will be a significant milestone and serve as a litmus test for new physics such as supersymmetry and the left-right symmetrical model. Furthermore, at the STCF experiment, the sensitivity of hyperon EDM measurements can be further enhanced by 2 orders of magnitude. Additionally, this angular analysis enables the determination of $CP$ violation in hyperon decays, the effective weak mixing angle, and beam polarization.

\end{abstract}



\maketitle

The measurement of a particle's permanent electric dipole moment (EDM), which violates both parity ($P$) and time-reversal symmetries, and, consequently, charge parity ($CP$) symmetry according to the $CPT$ theorem, serves as not only a robust means of testing the validity of the Standard Model (SM) but a sensitive probe for new physics, especially those that could induce lower-loop or flavor-diagonal $CP$ Violation (CPV), in the multi-100-TeV mass range~\cite{Beacham:2019nyx,Chupp:2017rkp}.
Neutron and $^{199}$Hg EDM measurement have set an upper limit on the SM QCD effective vacuum phase of $\bar\theta\lessapprox10^{-10}$, yet the SM permits any value within the $[0,2\pi]$ range.
This conundrum is commonly known as the strong $CP$ problem~\cite{Kim:2008hd}.
Examining EDM within the hadronic system serves as a means to either corroborate or disprove the $\bar{\theta}$ explanation and, in conjunction with the investigation of leptonic EDM,
constitutes an essential approach for the pursuit of new physics~\cite{Beacham:2019nyx}.
Investigating EDM in baryonic and light nuclear systems presents a unique prospect for elucidating various CPV models~\cite{Dekens:2014jka}.
Within the hyperon system, the strange quark may exhibit a special interaction with new physics, potentially leading to a substantial EDM effect.
This could suggest that the new physics possesses a specific flavor structure.
Another crucial aspect is that the sole measurement of EDM is inadequate to differentiate various sources of CPV beyond the SM. Hence, it becomes essential to utilize supplementary observations from diverse systems, such as hadrons, atoms, nuclei, and molecules~\cite{Chupp:2017rkp}.

Despite over seven decades of study on the pursuit of EDMs, the $\Lambda$ hyperon remains the only member of the hyperon family for which the upper limit of EDM, $1.5\times 10^{-16}$ $e$ cm, has been measured via spin precession at Fermilab~\cite{Pondrom:1981gu}.
The absolute value of the $\Lambda$ EDM, which is estimated indirectly using the experimental upper limit of the neutron EDM, is $< 4.4\times 10^{-26}$ $e$ cm~\cite{Guo:2012vf,Atwood:1992fb,Pich:1991fq,Borasoy:2000pq}.
There are no indirect predictions for hyperons with two or three strange valence quarks.
A variety of experimental approaches have been proposed, such as $\Lambda$ EDM measurement utilizing spin precession induced by dipole magnet at the LHCb experiment~\cite{Botella:2016ksl} and $\overline{\Xi}^+$ and $\overline{\Omega}^+$ EDM measurements employing spin precession induced by the strong field inside bent crystal at a fixed-target experiment~\cite{Bagli:2017foe}.
The direct measurement of EDM using spin precession is a challenging task due to the really limited lifetimes of hyperons.
Challenges also arise from the different production mechanisms and lifetimes of hyperons, making it difficult to prepare sources of various hyperons for EDM measurements in a single fixed-target experiment.

In contrast to fixed-target experiments and hadron collider experiments, a large number of entangled $\Lambda$, $\Sigma$, and $\Xi$ hyperon-antihyperon pairs can be readily produced and reconstructed from charmonium $J/\psi$ decays at tau-charm factories.
The considerable production cross section of $J/\psi$ in $e^+e^-$ annihilation, along with the large branching fraction of $J/\psi$ to hyperon-antihyperon pairs and the outstanding performance of contemporary detectors, guarantees that the identification of hyperon-antihyperon pairs is typically accomplished with a purity exceeding $95\%$.
This capability allows the exploration of tiny violations of conservation laws~\cite{BESIII:2020nme,Achasov:2023gey,Li:2016tlt}.
The production of entangled hyperon-antihyperon pairs, wherein the electric dipole form factor is incorporated within the $P$- and $CP$-violating components of the Lorentz invariant amplitude, offers a distinctive opportunity for indirectly extracting hyperon EDM. The electric dipole form factor is typically a complex number in the context of nonzero timelike momentum transfer, whereas it converges to the EDM in the limit of zero momentum transfer.
In practice, this kind of form factor can be effectively treated as an EDM on the assumption that the momentum transfer dependence is negligible due to an unknown extension to the zero region.

This paper reports a proposal to extract the hyperon EDMs through full angular analysis.
EDM measurements will be discussed in $e^+e^-$ collision within the region of $J/\psi$ resonance, considering two different types: {\bf(i)} $J/\psi\to B\overline{B}$ where $B$ are $\Lambda$ and $\Sigma^+$ hyperons; {\bf(ii)} $J/\psi\to B\overline{B}$ where $B$ are $\Xi^{-}$ and $\Xi^{0}$. Sequential hyperon decays are reconstructed by $\Lambda\to p\pi^-$, $\Sigma^+\to p\pi^0$, $\Xi^{-}\to\Lambda\pi^-$, and $\Xi^0\to\Lambda\pi^0$, correspondingly.
The utilization of multidimensional information in the complete decay chain significantly improves the sensitivity of the EDM measurement compared to one-dimensional analysis, such as a $CP$-odd triple-product moment encompassing hyperons $\Lambda$, $\Sigma^+$, $\Xi^-$, and $\Xi^0$~\cite{He:1992ng,He:1993ar,Zhang:2009at,Zhang:2010zzo,He:2022jjc}.
Scenarios for the BESIII experiment and a proposed future Super Tau-Charm Facility (STCF) are studied, of which the former one has already collected the world's largest dataset of 10 billion $J/\psi$ particles~\cite{BESIII:2021cxx}, and the latter one is devised to accumulate approximately $3.4\times10^{12}$ $J/\psi$ particles per year~\cite{Achasov:2023gey}.

\begin{figure}[ht!]
    \centering
    \subfigure{
    \includegraphics[width=0.95\linewidth]{"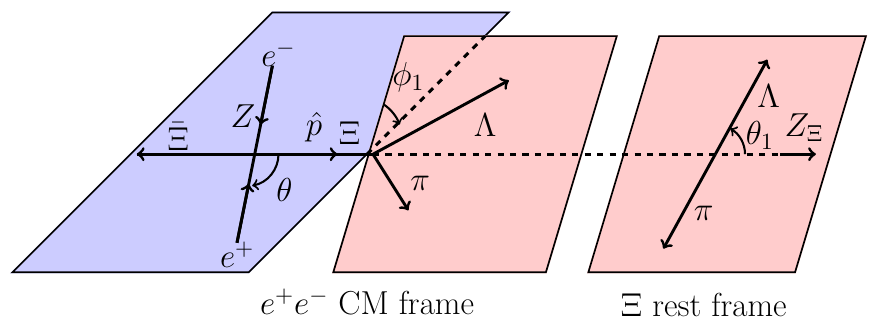"}}
    \subfigure{
    \includegraphics[width=0.95\linewidth]{"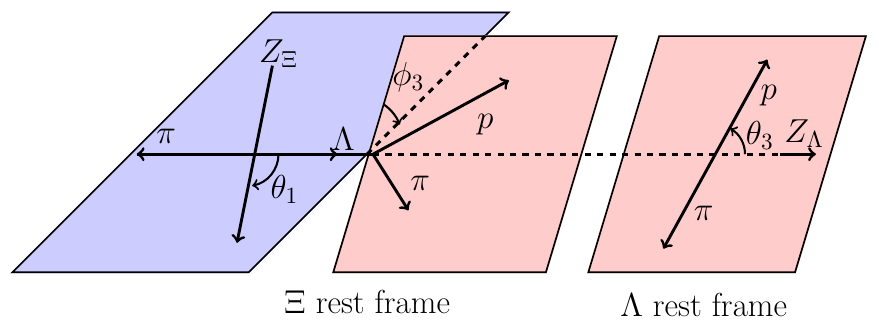"}}
    \caption{The Cartesian coordinate system of the $\Xi$ hyperon, denoted as ($X_{\Xi}$, $Y_{\Xi}$, $Z_{\Xi}$), is defined in the rest frame of $\Xi$, with the unit vector $Z_{\Xi}$ aligned with the momentum direction of $\Xi$, $\hat{p}$, and the unit vectors $Y_{\Xi}=Z\times Z_{\Xi}$ and $X_{\Xi}=Y_{\Xi}\times Z_{\Xi}$, respectively.
The motion direction of $\Lambda$ is characterized by the angles ($\theta_{1},\phi_{1}$).
The coordinate system of $\overline{\Xi}$ is defined by equalling  $(X_{\overline{\Xi}},Y_{\overline{\Xi}},Z_{\overline{\Xi}})$ to $(-X_{\Xi},Y_{\Xi},-Z_{\Xi})$.
The coordinate system of the $\Lambda$ hyperon, $(X_{\Lambda},Y_{\Lambda},Z_{\Lambda})$, is defined in the rest frame of $\Lambda$, with the unit vector $Z_{\Lambda}$ pointing to the direction 
of the $\Lambda$ momentum in the $\Xi$ frame, and the unit vectors $Y_{\Lambda}=Z_{\Xi}\times Z_{\Lambda}$ and $X_{\Lambda}=Y_{\Lambda}\times Z_{\Lambda}$, respectively. The proton moves along the direction of ($\theta_{3},\phi_{3}$).
Similarly, the coordinate system of $\overline{\Lambda}$ is determined based on that of $\bar{\Xi}$.}
    \label{fig:frame}
\end{figure}

Charmonium $J/\psi$ is produced via $e^+e^-$ annihilation, where the interference between the contributions from virtual $\gamma$ and $Z$-boson exchanges leads to a small longitudinal polarization of the $J/\psi$ meson.
The leading contribution from $Z$-boson exchange in SM, which violates parity symmetry, is suppressed by a factor of $M^2_{J/\psi}/m^2_Z$. Polarization effects are encoded in the spin density matrix of the $B\overline{B}$ hyperon pair, defined as
\begin{equation}
\begin{aligned}
R(\lambda_{1},\lambda_{2};\lambda^{\prime}_{1},\lambda^{\prime}_{2})\propto \sum_{m,m^{\prime}}&\rho_{m,m^{\prime}}d^{j=1}_{m,\lambda_{1}-\lambda_{2}}(\theta)d^{j=1}_{m^{\prime},\lambda^{\prime}_{1}-\lambda^{\prime}_{2}}(\theta)\\
&\times\mathcal{M}_{\lambda_{1},\lambda_{2}}\mathcal{M}^{*}_{\lambda^{\prime}_{1},\lambda^{\prime}_{2}}\delta_{m,m^{\prime}},
\end{aligned}
\end{equation}
where the indices $m^{(\prime)}$ and $\lambda^{(\prime)}_{1,2}$ represent the helicities of the $J/\psi$ meson and $B(\overline{B})$ hyperons, respectively.
The symbol $\rho_{m,m^{\prime}}$ is the spin density matrix of $J/\psi$ meson,
$d^{j}_{m^{(\prime)},\lambda^{(\prime)}_1-\lambda^{(\prime)}_2}(\theta)$ corresponds to the Wigner rotational function, and $\mathcal{M}_{\lambda^{(\prime)}_{1},\lambda^{(\prime)}_{2}}$ denotes the helicity amplitude of $J/\psi\to B\overline{B}$. The symbol $\theta$, as shown in Fig~\ref{fig:frame}, represents the angle between the momentum of the hyperon $B$, denoted as $\hat{p}$, and the motion of electron beam that is defined as the $Z$ axis.
The helicity states of the $J/\psi$ meson are represented by the helicity values $m^{(\prime)}$ of $+$, $-$, and $0$. The $3\times3$ matrix $\rho_{m,m^{\prime}}$ is reduced to a $2\times2$ matrix due to the suppression of the component $\rho_{00}$ by a factor of $m^2_e/M^2_{J/\psi}$.

The Lorentz invariant helicity amplitude in $J/\psi\to B\overline{B}$ decay with four independent form factors fixed at $q^2=M^2_{J/\psi}$ is written as~\cite{He:2022jjc} 
\begin{equation}
	\begin{aligned}
		\mathcal{M}_{\lambda_{1},\lambda_{2}}=\epsilon_{\mu}(\lambda_1-\lambda_2)\bar{u}(\lambda_{1},p_1) (F_{V}\gamma^{\mu}+\frac{i}{2m}\sigma^{\mu\nu}q_{\nu}H_{\sigma}\\
		+\gamma^{\mu}\gamma^{5}F_{A}+\sigma^{\mu\nu}\gamma^{5}q_{\nu}H_{T} )v(\lambda_{2},p_2),
	\end{aligned}
\end{equation}
where $m$ is $B$ hyperon mass, and $p_{1}$ and $p_{2}$ are the four momentum of $B$ and $\overline{B}$, respectively.

Processes involving a flavor-diagonal $CP$-violating vertex contribute to the electric dipole form factor $H_T$. The incorporation of $CP$-violating operators into a comprehensive Lagrangian is of paramount importance, as it serves as a vital connection between hyperon EDM and the underlying fundamental theories.
The various extensions of the SM give rise to unique contributions to these operators, leading to different impacts on the hyperon EDM.
Taking the $\Lambda$ hyperon as an example, multiple formulations have been presented to assess the influences stemming from the QCD $\theta$ term~\cite{Guo:2012vf,Pich:1991fq,Borasoy:2000pq}, quark chromoelectric dipole moment (qCEDM), four-quark operators~\cite{Faessler:2006at}, and the quark EDM (qEDM)~\cite{Anselm:1978vu}. Hyperon EDM measurements provide a direct means to evaluate the contributions from qEDM and qCEDM, owing to the suppressed effects of high-dimensional operators and the experimental constraint imposed by neutron EDM measurements on the QCD $\theta$ term.
The flavor-diagonal $CP$-violating contributions in the SM are extremely tiny, while new physics, such as supersymmetry (SUSY) and the left-right symmetrical model, may significantly enhance hyperon EDM as discussed extensively by analyzing EDM results from electron, neutron, and $^{199}$Hg systems~\cite{Chupp:2017rkp}.
Any unexpectedly large hyperon EDM will suggest a special coupling between the strange quark and new physics.
Consequently, in the decay chain under consideration, we provide an opportunity to explore these potential impacts within the hyperon family by establishing a connection between $H_{T}$ and the contribution of hyperon EDM~\cite{He:1992ng,He:1993ar,He:2022jjc} as
\begin{equation}
	\begin{aligned}
        H_{T}=\frac{2e}{3M^{2}_{J/\psi}}g_{V}d_{B}.
	\end{aligned}
\end{equation}
The form factor $H_{T}$ here, in fact, varies with $q^{2}$. Neglecting the reliance on $q^{2}$, $d_{B}$ is then the EDM of hyperon $B$. Considering the dispersive part of timelike reaction, the imaginary part of $H_{T}$ is also investigated in this angular analysis.
The aforementioned discussions will also be applied to the hyperons $\Sigma$ and $\Xi$ in this paper.

The form factors $F_{V}$ and $H_{\sigma}$ are related to the redefined $G_{1,2}$ as described in~\cite{He:2022jjc}
\begin{equation}
	\begin{aligned}
		 F_{V}=G_{1}-4m^{2}\frac{(G_{1}-G_{2})}{(p_{1}-p_{2})^{2}},~~H_{\sigma}=4m^{2}\frac{(G_{1}-G_{2})}{(p_{1}-p_{2})^{2}}.
	\end{aligned}
\end{equation}
The form factors $G_{1}$ and $G_{2}$ are associated to the experimental observables $\alpha_{J/\psi}$, $\Delta\Phi$, and $\Gamma(J/\psi\to B\overline{B})$ through the relationship $\alpha_{J/\psi}=\frac{M^{2}{J/\psi}\left|G{1}\right|^{2}-4m^{2}\left|G_{2}\right|^{2}}{M^{2}{J/\psi}\left|G{1}\right|^{2}+4m^{2}\left|G_{2}\right|^{2}}$ and $\frac{G_{1}}{G_{2}}=\left|\frac{G_{1}}{G_{2}}\right|e^{-i\Delta\Phi}$~\cite{BESIII:2018cnd,BESIII:2022qax}.
The form factor $F_A$, primarily arising from $Z$-boson exchange between $c\overline{c}$ and light quark pairs $q\overline{q}$ within the SM, can be related to the effective weak mixing angle $\theta^{\text{eff}}_W$ through
\begin{equation}
	\begin{aligned}
        F_{A}\approx -\frac{1}{6}Dg_{V}\frac{g^{2}}{4\cos^2\theta^{\text{eff}}_W}\frac{1-8\sin^2\theta^{\text{eff}}_W/3}{m^{2}_{Z}},
	\end{aligned}
\end{equation}
which leads to a parity violation effect estimated to be on the order of $10^{-6}$. The symbol $g_{V}$ is defined as $\bra{0}\bar{c}\gamma^{\mu}c\ket{J/\psi}=g_{V}\epsilon^{\mu}$, and $D$ is a nonperturbative parameter that is fitted from data~\cite{He:2022jjc}. Precise measurements utilizing large statistics make it feasible to extract the weak mixing angle $\sin^2\theta^{\text{eff}}_W$, which is essential in testing the SM, particularly in regards to the effects derived from quantum corrections of heavy particles, such as the Higgs boson and the top quark, at the loop level~\cite{Kumar:2013yoa}.

The longitudinal polarization of the $J/\psi$ meson, denoted as $P_L$, is defined as the relative difference between the diagonal elements of the density matrix, $\rho_{++}$ and $\rho_{--}$. Moreover, in experiments such as BESIII where there is no beam polarization, the polarization $P_L$ is closely connected to the left-right asymmetry $\mathcal{A}^{0}_{LR}$, 
\begin{equation}
	\begin{aligned}
	 P_L=\mathcal{A}^{0}_{LR}=\frac{\sigma_{R}-\sigma_{L}}{\sigma_{R}+\sigma_{L}}=\frac{-\sin^{2}\theta^{\text{eff}}_{W}+3/8}{2\sin^{2}\theta^{\text{eff}}_{W}\cos^{2}\theta^{\text{eff}}_{W}}\frac{M^{2}_{J/\psi}}{m^{2}_{Z}}.
	\end{aligned}
\end{equation}
Here, $\sigma_{R(L)}$ represents the $J/\psi$ cross section with right-handed (left-handed) electrons. This asymmetry is induced by the effective weak mixing angle $\theta^{\text{eff}}_{W}$ and, hence, suppressed to the order of $10^{-4}$~\cite{Bondar:2019zgm}.
For experiments with an electron beam longitudinally polarized at the magnitude of $P_e$, like STCF~\cite{Achasov:2023gey}, the $P_L$ can be replaced by $\xi$:
\begin{equation}
	\begin{aligned}
	 \xi=\frac{\sigma_{R}(1+P_{e})/2-\sigma_{L}(1-P_{e})/2}{\sigma_{R}(1+P_{e})/2+\sigma_{L}(1-P_{e})/2}=\frac{\mathcal{A}^{0}_{LR}+P_{e}}{1+P_{e}\mathcal{A}^{0}_{LR}}\approx P_{e}.
	\end{aligned}
\end{equation}
In this case, the longitudinal polarization of the electron beam instead of $Z$-boson exchange will play a crucial role in enhancing the sensitivity of measurements.

Based on the rotational symmetry, the helicity representation of the complete angular distribution for type {\bf(ii)} is given as
\begin{equation}\label{angdis}
\begin{aligned}
 \frac{d\sigma}{d\Omega}\propto \sum_{\left[\lambda\right]}&R(\lambda_{1},\lambda_{2};\lambda^{\prime}_{1},\lambda^{\prime}_{2})\\
 &D^{*j=1/2}_{\lambda_{1},\lambda_{3}}(\phi_{1},\theta_{1})D^{j=1/2}_{\lambda^{\prime}_{1},\lambda^{\prime}_{3}}(\phi_{1},\theta_{1})\mathcal{H}_{\lambda_{3}}\mathcal{H}^{*}_{\lambda^{\prime}_{3}}\\
 &D^{*j=1/2}_{\lambda_{2},\lambda_{4}}(\phi_{2},\theta_{2})D^{j=1/2}_{\lambda^{\prime}_{2},\lambda^{\prime}_{4}}(\phi_{2},\theta_{2})\overline{\mathcal{H}}_{\lambda_{4}}\overline{\mathcal{H}}^{*}_{\lambda^{\prime}_{4}}\\
 &D^{*j=1/2}_{\lambda_{3},\lambda_{5}}(\phi_{3},\theta_{3})D^{j=1/2}_{\lambda^{\prime}_{3},\lambda_{5}}(\phi_{3},\theta_{3})\mathcal{F}_{\lambda_{5}}\mathcal{F}^{*}_{\lambda_{5}}\\
 &D^{*j=1/2}_{\lambda_{4},\lambda_{6}}(\phi_{4},\theta_{4})D^{j=1/2}_{\lambda^{\prime}_{4},\lambda_{6}}(\phi_{4},\theta_{4})\overline{\mathcal{F}}_{\lambda_{6}}\overline{\mathcal{F}}^{*}_{\lambda_{6}}\\
\end{aligned}
\end{equation}
where $\left[\lambda\right]$ is a set containing all of the possible helicity symbols appearing in the summation like $\lambda_{1},\lambda_{2},\lambda^{\prime}_{1},\lambda^{\prime}_{2}....$
The momenta directions of $\Lambda$ and $\overline{\Lambda}$ in the frame of $\Xi$ and $\overline{\Xi}$ are parametrized by the polar and azimuthal angles $\theta_{1},\phi_{1}$ and $\theta_{2},\phi_{2}$, respectively.
The polar and azimuthal angles $\theta_{3},\phi_{3}$ and $\theta_{4},\phi_{4}$ refer to the orientation of the proton and antiproton in the frame of $\Lambda$ pairs, respectively.
The definitions of these helicity angles are illustrated in Fig~\ref{fig:frame}, and analogous definitions are employed for the subsequential decay of antiparticles.
Helicity amplitudes $\mathcal{H}_{\lambda_{i}}$ and $\mathcal{F}_{\lambda_{i}}$ are used to parametrize the dynamics of weak decay $\Xi\to \Lambda\pi$ and $\Lambda\to p\pi$,
and corresponding charge conjugated processes are denoted by $\mathcal{H}$ and $\mathcal{F}$ with a bar. The formula for type {\bf(i)} is obtained by retaining only $\theta_{1,2}$ and $\phi_{1,2}$ and identifying $\mathcal{H}$ as $\mathcal{F}$.

The asymmetry parameters $\alpha$ and $\phi$, which were originally introduced by Lee and Yang~\cite{Lee:1957qs}, are used to define the hyperon $CP$-violating observables as $A^{B}_{CP}=(\alpha_{B}+\bar{\alpha}_{B})/(\alpha_{B}-\bar{\alpha}_{B})$ and $\Delta\phi^{B}_{CP}=(\phi_{B}+\bar{\phi}_{B})/2$~\cite{BESIII:2021ypr}.
Two observables are complementary as they rely on the sine and cosine of the strong phase difference, respectively. In hyperon decays, the relative strong phases are small, so that the latter one exhibits better sensitivity~\cite{Wang:2022fih,Donoghue:1986hh}. Moreover, in this paper, the latter in $\Xi$ decays can be determined because of the measurable polarization of the $\Lambda$ hyperon.

\begin{table}[!hb]
\centering
\caption{Estimated yields of pseudoexperiments based on the statistics from BESIII and STCF experiments, where $B_{tag}$ represents the branching ratio of cascade decay, $\epsilon_{tag}$ represents the expected detection efficiency, and $N_{tag}^{evt}$ represents the number of expected events after reconstruction.
\label{tab:input_para}}
\scalebox{0.7}{
    \begin{tabular}{l|c|c|c|c}
    \hline
    \hline
    Decay channel & $J/\psi \rightarrow \Lambda\bar{\Lambda}$ & $J/\psi \rightarrow \Sigma^{+}\bar{\Sigma}^{-}$ & $J/\psi \rightarrow \Xi^{-}\bar{\Xi}^{+}$ & $J/\psi \rightarrow \Xi^{0}\bar{\Xi}^{0}$ \\
    \hline
    $B_{tag}/(\times 10^{-4})$~\cite{ParticleDataGroup:2022pth} & 7.77 & 2.78 & 3.98 & 4.65 \\
    \hline
    $\epsilon_{tag}/\%$~\cite{BESIII:2022qax,BESIII:2020fqg,BESIII:2021ypr,BESIII:2023drj} & 40 & 25 & 15 & 7 \\
    \hline
    $N_{tag}^{evt}/(\times 10^{5})$(BESIII) & 31.3 & 7.0 & 6.0 & 3.3 \\
    \hline
    $N_{tag}^{evt}/(\times 10^{8})$(STCF)~\cite{Achasov:2023gey} & 10.6 & 2.4 & 2.0 & 1.1 \\
    \hline
    \hline
    \end{tabular}
}
\end{table}

\begin{figure*}[!t]
    \centering
    \subfigure[Sensitivity for $Re(d_{B})$ and $Im(d_{B})$]{
    {\label{fig:sensitivity_a}}
    \includegraphics[width=0.45\linewidth]{"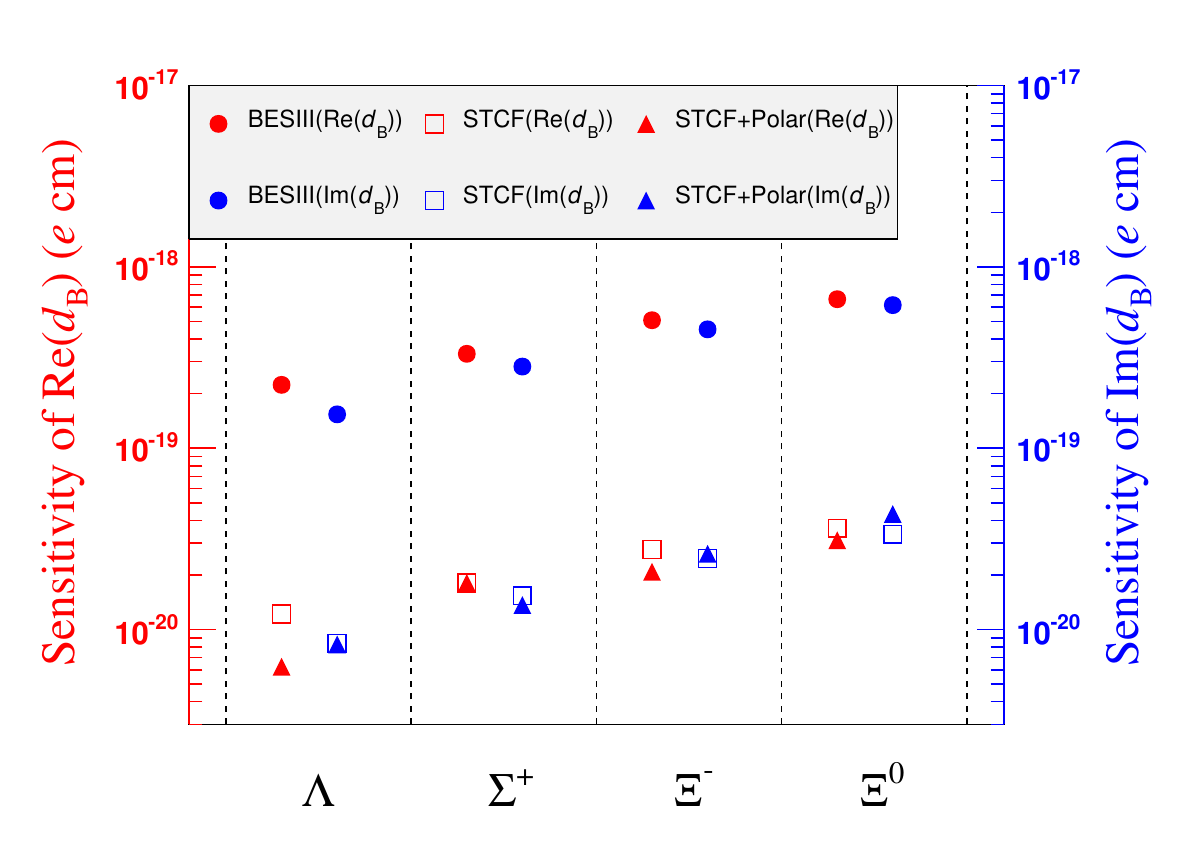"}}
    \subfigure[Sensitivity for $A_{CP}^{B}$ and $\Delta\phi_{CP}^{B}$]{
    {\label{fig:sensitivity_b}}
    \includegraphics[width=0.45\linewidth]{"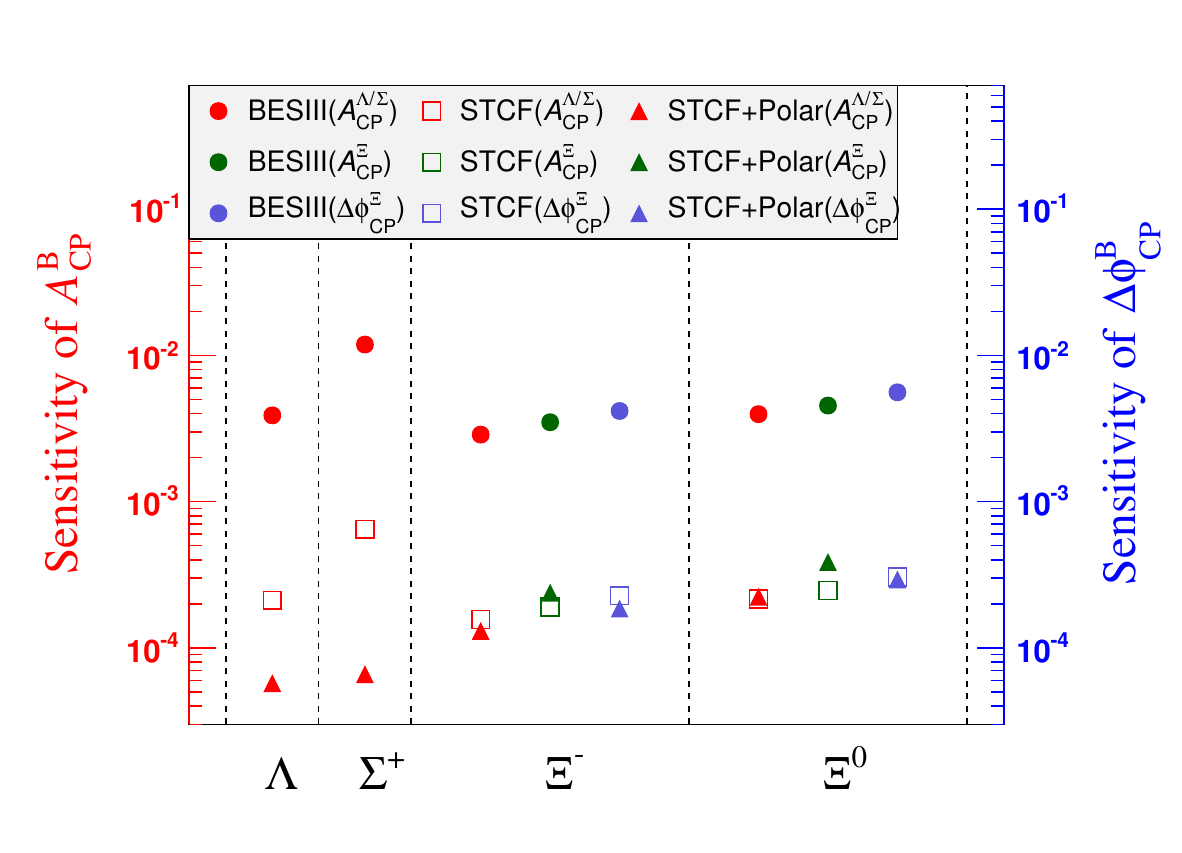"}}    
    \subfigure[Sensitivity for $|F_{A}|$ and $\text{sin}^{2}\theta_{W}^{\text{eff}}$]{
    {\label{fig:sensitivity_c}}
    \includegraphics[width=0.45\linewidth]{"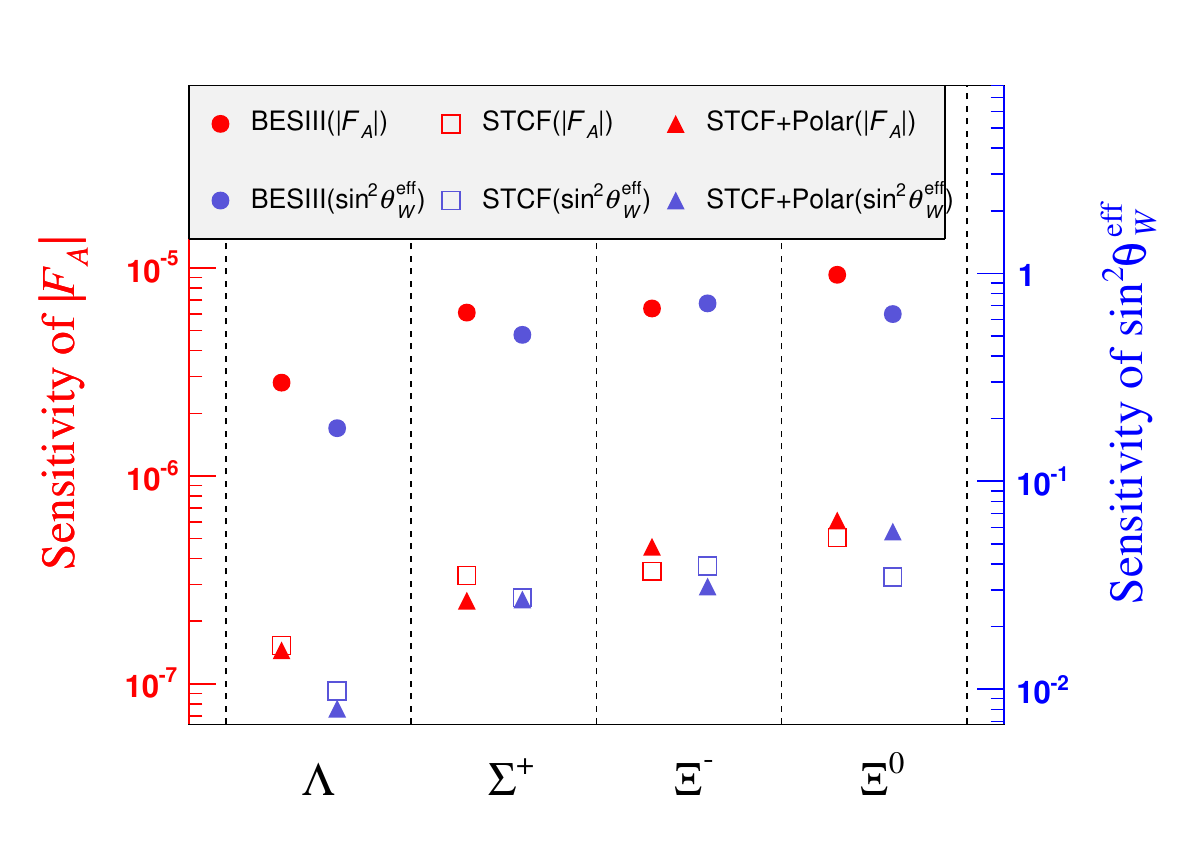"}}
    \subfigure[Sensitivity for $P_{L}$ and $\text{sin}^{2}\theta_{W}^{\text{eff}}$]{
    {\label{fig:sensitivity_d}}
    \includegraphics[width=0.45\linewidth]{"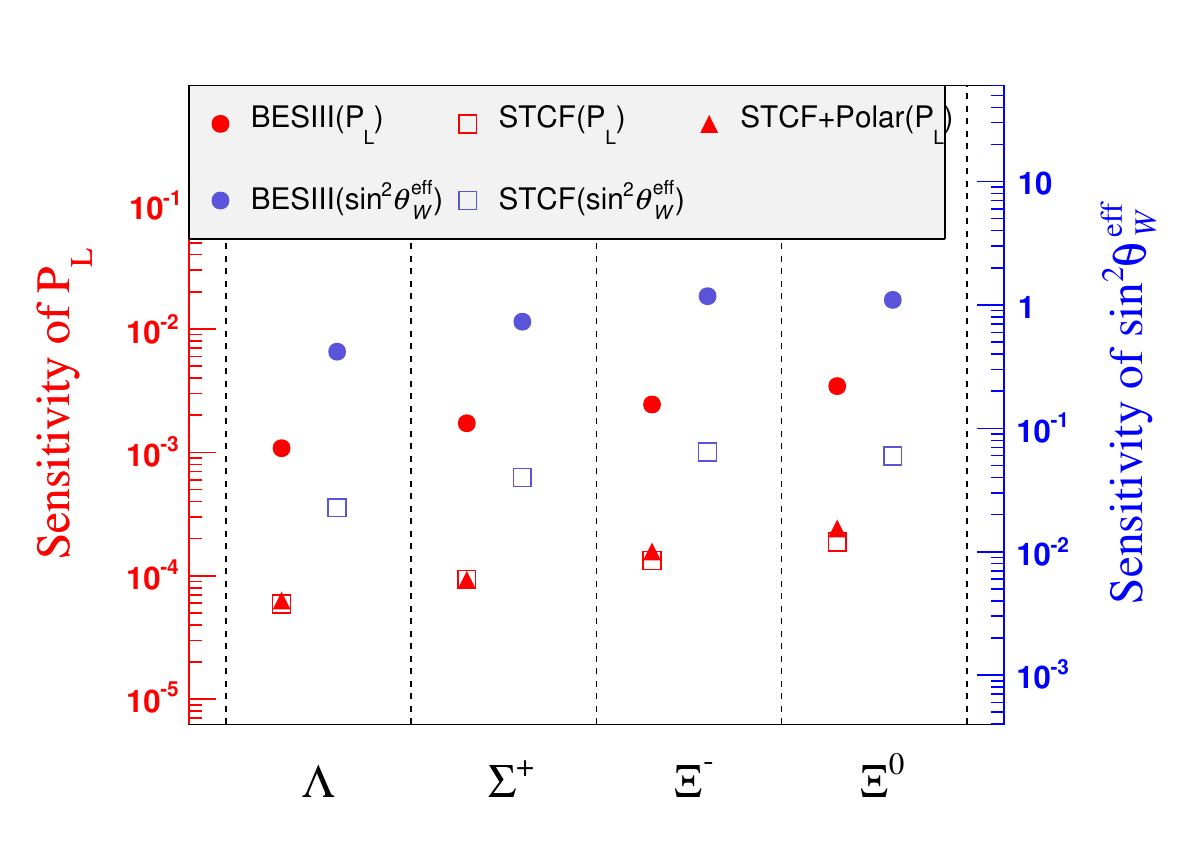"}}
        \caption{\label{fig:sensitivity}
	Sensitivity for (a) EDM, (b) $CP$-violating variables in hyperon decays, and (c) form factor $F_A$, (d) polarization for $J/\psi$ product and beam, and (c), (d) weak mixing angle parameter $\text{sin}^{2}\theta_{W}^{\text{eff}}$. The markers for hyperons $\Lambda$, $\Sigma^{+}$, $\Xi^{-}$, and $\Xi^{0}$ are located within dashed regions, respectively. The red and blue markers correspond to the sensitivity of the physical quantities represented by the red title (left Y axis) and blue title (right Y axis) in the graph. The green markers in (b) correspond to the physical quantity represented by the red title on the left Y axis. The full circle, open square, and full triangle up correspond to the estimated sensitivities for the BESIII experiment, the STCF experiment with an unpolarized beam, and the STCF experiment with $80\%$ polarized electron beam, respectively.}
\end{figure*}

To evaluate the statistical sensitivity of the measurement, 500 pseudoexperiments of each decay are generated and fitted using a probability density function derived from the full angular distributions shown in Eq.~\eqref{angdis}.
The estimated yields presented in Table~\ref{tab:input_para},
as well as the form factors and decay parameters from the published articles~\cite{He:2022jjc, ParticleDataGroup:2022pth, BESIII:2022qax,BESIII:2020fqg,BESIII:2021ypr,BESIII:2023drj}, are fixed for the generation.
The EDM, along with other form factors, decay parameters, and polarization, can be simultaneously determined from fitting.
This work further investigates the sensitivities for different statistics at BESIII and STCF experiments, taking into account branching fractions, detection efficiencies, and the impact of a longitudinally polarized electron beam.
Figure~\ref{fig:sensitivity_a} presents the estimated sensitivities for hyperon EDMs.
Utilizing the present statistics from the BESIII experiment, the sensitivity for $\Lambda$ EDM reaches $10^{-19}$ $e$ cm (red full circle), achieving a remarkable 3-orders-of-magnitude improvement over the only existing measurement at Fermilab with similar statistics. Furthermore, the BESIII experiment maintains state-of-the-art sensitivities of $10^{-19}$ $e$ cm for the $\Sigma^+$, $\Xi^-$, and $\Xi^0$ hyperons. The STCF experiment is expected to further improve the EDM sensitivities by around 1-2 orders of magnitude (shown by the open square and full triangle symbols).

The calculated sensitivities for CPV in hyperon decays are depicted in Figure~\ref{fig:sensitivity_b}.
With an $80\%$ longitudinally polarized electron beam at STCF experiment, the most optimal sensitivities for CPV induced by the $\alpha_B$ parameter (red full triangle) are expected to reach $5\times10^{-5}$ ($6\times10^{-5}$) in $J/\psi\to\Lambda\overline{\Lambda}$ ($J/\psi\to\Sigma^+\overline{\Sigma}^-$) decays. Similarly, for the $\phi_B$ parameter, the sensitivities for CPV in the decays $J/\psi\to\Xi^-\overline{\Xi}^+$ and $J/\psi\to\Xi^0\overline{\Xi}^0$ are $2\times10^{-4}$ and $3\times10^{-4}$ respectively, as represented by the blue full triangle.
The sensitivities for the observables $A^{B}_{CP}$ and $\Delta\phi^{B}_{CP}$ have achieved the anticipated levels as predicted by the SM~\cite{Deshpande:1994vp,Tandean:2002vy,Donoghue:1986hh}.
The estimated sensitivities for $F_A$ and the subsequently constrained $\sin^2\theta^{\text{eff}}_W$ are depicted in Figure~\ref{fig:sensitivity_c}, where only the sensitivities for the module of $F_A$ are reported as the phase is found to have a small impact based on the toy study.
The sensitivity for $\sin^2\theta^{\text{eff}}_W$ constrained solely by $F_A$ can reach $8\times10^{-3}$.
Similarly, Figure~\ref{fig:sensitivity_d} depicts the estimated sensitivities for $J/\psi$ polarization and the subsequently constrained $\sin^2\theta^{\text{eff}}_W$ and, accordingly, the sensitivity for $\sin^2\theta^{\text{eff}}_W$ constrained solely by $P_L$ can reach $2\times10^{-2}$ at the STCF experiment. Furthermore, the sensitivity for $\sin^2\theta^{\text{eff}}_W$ will be further improved to $5\times10^{-3}$ in $J/\psi\to\Lambda\overline{\Lambda}$ decays by the simultaneous constraint from $F_A$ and $P_L$.
Longitudinal polarization for the electron beam can also be determined through angular analysis, yielding the best sensitivity up to $6\times10^{-5}$ as depicted in Figure~\ref{fig:sensitivity_d} (red full triangle up), which can be used for more precise measurement of the weak mixing angle from Bhabha scattering events~\cite{Bondar:2019zgm}.

In conclusion, to investigate the largely unexplored territory of hyperon EDMs, we have proposed a comprehensive angular analysis, considering $P$ violation in $J/\psi$ production and $CP$ and $P$ violation in $J/\psi$ decay. The EDM, along with $CP$-violating observables in hyperon decays, effective weak mixing angle, and beam polarization can be simultaneously extracted from the angular analysis. The statistical sensitivities for physical observables have been estimated for BESIII and STCF scenarios. Utilizing the expected statistics obtained from the BESIII experiment, the $\Lambda$ EDM measurement can achieve an impressive upper limit of $10^{-19}$ $e$ cm, presenting a remarkable improvement of 3 orders of magnitude compared to the only existing measurement at Fermilab with similar statistics. The EDM measurement of $\Sigma^+$, $\Xi^-$, and $\Xi^0$ hyperons at the same level of $10^{-19}$ $e$ cm will be a ground-breaking accomplishment as the first-ever achievement and the latter two will be the first exploration of hyperons with two strange valence quarks.
At the STCF experiment, with a longitudinally polarized electron beam, the search for hyperon EDMs could potentially reach levels of $10^{-21}-10^{-20}$ $e$ cm.
The EDM measurements for hyperons will be a significant milestone and a stringent test for new physics, such as SUSY and the left-right symmetrical model.
At the same time, the CPV in hyperon decays could be verified at a sensitivity of $10^{-5}-10^{-4}$, which has already matched the predictions of the SM.
The effective weak mixing angle parameter can be measured at a level of $10^{-3}$ and further enhanced by utilizing the precisely determined beam polarization in the angular analysis.
This method can also be extended to $\psi(2S)$ decays for investigating the pure strange quark hyperon $\Omega$ with taking into account additional form factors arising from its spin-$\frac{3}{2}$ property.


\begin{acknowledgments}
We thank Professor Fengkun Guo, Professor Xiaogang He, Professor Jianping Ma, and Professor Yangheng Zheng for very useful discussion.
This work is supported by National Key R$\&$D Program of China No. 2022YFA1602204; National Natural Science Foundation of China (NSFC) under Contracts No. No. 11935018, No. 12221005, No. 11975112, and No. 12335003; Fundamental Research Funds for the Central Universities, University of Chinese Academy of Sciences, Lanzhou University (lzujbky-2023-it12).
\end{acknowledgments}

\bibliography{edmhyperon}
\end{document}